\documentclass[preprint,12pt]{elsarticle}

\usepackage{epsfig}
\usepackage{enumerate}
\usepackage[normalem]{ulem}
\usepackage{amssymb,amsmath,graphics,graphicx}  
\usepackage{xcolor}
\usepackage{mathtools}

\usepackage{booktabs} 
\usepackage{multirow}

\journal{Elsevier}

\begin{document}

\begin{frontmatter}




\title{The propensity for disobedience: Rule-breaking, compliance and social phase transitions}



\author{Nuno Crokidakis}
\ead{nunocrokidakis@id.uff.br}

\address{
Instituto de F\'{\i}sica, Universidade Federal Fluminense, Niter\'oi, Rio de Janeiro, Brazil 
}

\begin{abstract}
  We develop a mathematical model to describe the persistence of rule-breaking behaviors in societies, such as traffic violations, disregard for legal restrictions and other forms of noncompliance. Using a replicator-type dynamics with utility functions incorporating individual benefits, institutional punishment and social sanctions, we first built a general formulation of the system. Within this framework, we analyze two distinct models differing in the nature of social feedback. In the presence of positive feedback, the system exhibits bistability, with widespread compliance and widespread violation as stable equilibria, and the transition between these states occurs discontinuously once a critical threshold is crossed, resembling a first-order phase transition. By contrast, when negative feedback is present, the population undergoes a continuous phase transition between compliant and noncompliant collective states, driven by an increasing collective cost of rule-breaking. Numerical simulations and analytical results illustrate how changes in enforcement, social tolerance or perceived benefits can shift the system across critical thresholds separating distinct collective regimes. The results provide a theoretical explanation for the fragility of social order under weak institutions and highlight possible pathways to promote compliance.
\end{abstract}

\begin{keyword}
Dynamics of social systems \sep Collective phenomena \sep Evolutionary dynamics \sep Social norms \sep Nonequilibrium phase transitions


\end{keyword}

\end{frontmatter}


\section{Introduction}

\quad Rule compliance and rule-breaking are pervasive features of human societies. From traffic violations and tax evasion to academic misconduct and informal
norm transgressions, individuals repeatedly face the decision between adhering
to established rules or pursuing private advantages \cite{Krause}. Although such choices may appear purely individual, their aggregate outcomes often display collective patterns, including abrupt norm collapse, persistent intermediate levels of compliance, or long periods of apparent stability followed by sudden shifts. In this sense, rule-breaking behavior can be viewed as an emergent phenomenon arising from interactions among individuals, rather than as the result of isolated decisions.

The dynamics of social norms have been studied across multiple disciplines using a variety of theoretical frameworks. Threshold models of collective behavior, such as those proposed by Granovetter \cite{Granovetter}, emphasize how individual decisions depend on the behavior of others, leading to nonlinear aggregate outcomes. Similarly, classical segregation models \cite{Schelling} illustrate how simple local incentives can generate large-scale macroscopic patterns.

From an economic and evolutionary perspective, several works have investigated the emergence and stability of social norms through game-theoretic and adaptive frameworks \cite{Young}. In parallel, statistical physics approaches have provided a powerful set of tools to study collective behavior as an emergent property of interacting agents \cite{pmco,galam_review,galam_book,sen_book,galam_frontiers,bar,Battiston}, with replicator dynamics and related models offering a natural description of strategy evolution in populations  \cite{Nowak,hofbauer,rmp}.

More recently, collective behavioral changes have also been studied in the context of cascading processes on complex networks, highlighting how local interactions can lead to large-scale shifts in social behavior \cite{Borge-Holthoefer}.

A key ingredient in many models of norm compliance is the role of punishment, social sanctions and reputational incentives. Experimental evidences suggest that enforcement mechanisms and peer sanctions can substantially influence cooperative behavior and norm adherence \cite{Fehr}, and external conditions such as collective risk can significantly influence the strength of social norms and promote cooperative behavior in groups \cite{Antonioni}.

However, less attention has been given to the structural distinction between different types of collective feedback and their impact on the macroscopic nature of social transitions \cite{pmco,galam_review,galam_book,sen_book,galam_frontiers}.

In particular, it remains unclear under which conditions the breakdown of social norms occurs abruptly, resembling a discontinuous phase transition, and when it instead proceeds gradually, leading to continuous changes in collective behavior. Addressing this question requires a minimal dynamical framework capable of distinguishing between reinforcing and stabilizing feedback mechanisms.

Most existing works focus either on equilibrium game-theoretic approaches or on empirical norm dynamics. Less attention has been given to the structural distinction between different types of collective feedback and its implications for the qualitative nature of social transitions. While similar dynamical frameworks have been widely used in the study of social norms and evolutionary dynamics, the present work focuses on this structural aspect of social interactions. In this work, we develop a mathematical model for the evolution of rule-breaking behavior based on adaptive utility comparison.

We first introduce a general formulation in which individual incentives, institutional enforcement and social feedback jointly determine strategy adaptation. We then analyze two distinct regimes: one characterized by positive social reinforcement, leading to bistability and discontinuous transitions, and another characterized by increasing collective costs of widespread violations, resulting in continuous transitions and intermediate stationary states.

In particular, we show that the qualitative nature of collective outcomes is not determined by individual rationality per se, but by the structural form of collective feedback. This provides a unified perspective linking the structure of social interactions to the order of social phase transitions.


\section{Models and Results}

\subsection{General model formulation}

We consider a population in which individuals repeatedly face the choice between complying with a social rule or breaking it. Let $d(t)\in[0,1]$ denote the fraction of rule-breakers at time $t$, so that $1-d(t)$ is the fraction of compliant individuals. The key idea is that the perceived advantages of each strategy depend not only on 
institutional enforcement, but also on collective behavior. We assume that agents choose between compliance (C) and disobedience (D) by comparing the corresponding expected utilities, which combine private incentives, institutional enforcement and social feedback.

In other words, we will define utility functions $U_C(d)$ and $U_D(d)$, for compliance and disobedience, respectively. Thus, the utility difference can be defined as 
\begin{equation}  \label{eq1}
\Delta(d)=U_D(d) - U_C(d) ~,      
\end{equation}
\noindent
and it can be viewed as a relative advantage of rule-breaking.

All quantities entering the utility functions are assumed to have the same effective dimension, corresponding to a payoff or utility. As a consequence, the dynamics depends only on differences in utilities, and the model can be expressed in dimensionless form by an appropriate rescaling of the parameters.

As we will show, the nature of the phase transitions observed in the model, whether continuous or discontinuous, is not a property of individual agents, but rather of the collective feedback mechanisms governing their interactions.

In the following subsections, we introduce the two distinct models analyzed in this work.


\subsection{Model I: Positive social feedback and bistable dynamics}

The expected utility of breaking the rule in Model I is written as
\begin{equation} \label{eq2}
U_D(d) = B - pF - s\,(1-d),
\end{equation}
where 
$B$ represents the immediate private benefit of disobedience, $p$ is the probability of punishment, $F$ is the fine (or institutional cost) and $s(1-d)$ captures social sanctions that become stronger when few people disobey.

The expected utility of compliance is
\begin{equation} \label{eq3}
U_C(d) = -C + R\,(1-d),
\end{equation}
where $C$ represents the individual cost of complying (time, effort, discomfort). The term $R\,(1-d)$ accounts for reputational or symbolic rewards associated with complying with the rule. These rewards depend on the social environment and tend to be stronger when rule-following behavior is widespread in the population, reflecting the social value attached to conformity.

From Eq. \eqref{eq1}, we have the utility difference for Model I,
\begin{equation} \label{eq4}
\Delta(d) = U_D(d) - U_C(d) = (B - pF + C) - (s+R)(1-d) = K - L(1-d),
\end{equation}
where
\begin{eqnarray} \label{eq5a}
K & = & B - pF + C  \\  \label{eq5b}
L & = & s + R
\end{eqnarray}

It is worth noting that all parameters entering the utility functions share the same effective dimension, so that the dynamics depends only on ratios between them. In particular, the relevant control parameters can be expressed in terms of dimensionless combinations such as $(B+C)/F$ and $(s+R)/F$.

We assume a replicator-type mean-field dynamics \cite{Nowak,hofbauer},
\begin{equation} \label{eq6}
\dot d(t) = \gamma\, d(1-d)\,\Delta(d),
\end{equation}
where $\gamma>0$ controls the speed of adaptation. The prefactor $d(1-d)$ ensures that $\dot d = 0$ at the absorbing boundaries $d=0$ (full compliance) and $d=1$ (full violation). We assume that individuals do not deterministically choose between compliance and disobedience, but rather adapt their behavior over time by comparing the relative utilities of the two strategies. In this context, the fraction $d(t)$ of disobedient agents evolves according to the difference in expected utilities, a standard assumption in evolutionary game theory and population dynamics \cite{hofbauer,szabo}. This leads naturally to a replicator-type equation for $d(t)$, ensuring that strategies with higher relative utility tend to increase in frequency. More specifically, the change in $d(t)$ is assumed to be proportional to both the fraction of agents currently disobeying and the fraction complying, reflecting the possibility of switching between strategies. This consideration leads to a logistic prefactor $d(1-d)$, while the direction and strength of the change are determined by the utility difference $\Delta(d)$. In this framework, the fraction of rule-breakers $d$ plays the role of an order parameter, whose stationary value $d^*$ determines the macroscopic phase of the system.

Besides $d=0$ and $d=1$, an internal fixed point exists whenever $0 < K < L$, in which case
\begin{equation} \label{eq7}
d^* = 1 - \frac{K}{L} = 1 - \frac{B - pF + C}{s + R} ~.
\end{equation}
\noindent
Linear stability analysis shows that $d^*$ is unstable, whereas the boundary states are stable. Thus, for $0<K<L$, the system is bistable: initial conditions below $d^*$ evolve toward a high-compliance phase ($d\rightarrow 0$), whereas initial conditions above $d^*$ evolve toward a high-violation phase ($d\rightarrow 1$). The point $d^*$ therefore acts as a critical threshold separating the basins of attraction. When parameters change such that $K> L$ or $K< 0$, the internal fixed point leaves the physical interval $[0,1]$, and only one absorbing phase remains. This mechanism produces a discontinuous transition between the two macroscopic regimes $d=0$ and $d=1$. Summarizing:
\begin{itemize}
\item $K>0$:
\begin{itemize}
\item $K>L$: $d^*<0$ and $\Delta(d)>0$, which implies that the relative advantage of rule-breaking is positive, for all $d\in [0,1]$, and the population converges towards $d=1$ (full violation) 

\item $0<K<L$: $d^*>0$ and $\Delta(d)$ can be positive or negative, depending on the initial condition for $d(t)$ (above or below $d^*)\,\Rightarrow$ bistability
\end{itemize}

\item $K<0$:  $d^*>1$ and $\Delta(d)<0$, which implies that the relative advantage of rule-breaking is negative, for all $d\in [0,1]$, and the population converges towards $d=0$ (full compliance) 
    
\end{itemize}

From a statistical physics perspective, this behavior reflects a generic mechanism in complex systems: macroscopic collective states are selected by dynamical stability rather than by individual-level considerations. Similar bistable structures and discontinuous transitions appear in a wide range of contexts, including coordination models, evolutionary games, magnetic systems and coupled epidemic-opinion dynamics, highlighting the universality of the underlying mechanism \cite{pmco,galam_review,galam_book,sen_book,rmp,andre_marcelo}.

In this formulation, $B$ and $C$ encode private incentives, $pF$ represents institutional enforcement, and $s$ and $R$ capture social feedback mechanisms. The model shows that even small individual advantages for rule-breaking may lead to large-scale shifts in collective behavior once a critical threshold is crossed, emphasizing the intrinsically collective nature of rule compliance.

\begin{figure}[t]
\begin{center}
\vspace{6mm}
\includegraphics[width=0.35\textwidth,angle=0]{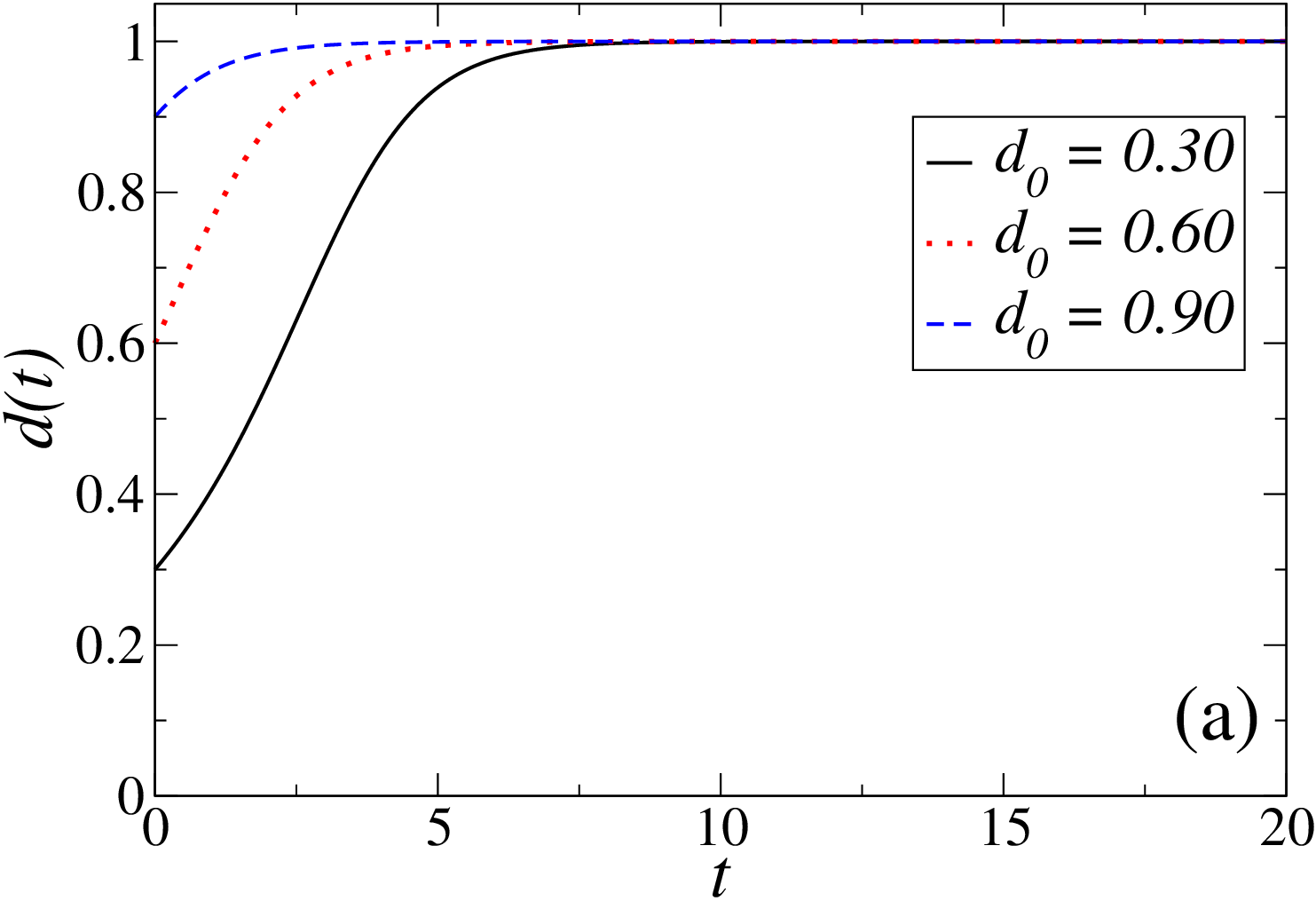}
\hspace{0.3cm}
\includegraphics[width=0.35\textwidth,angle=0]{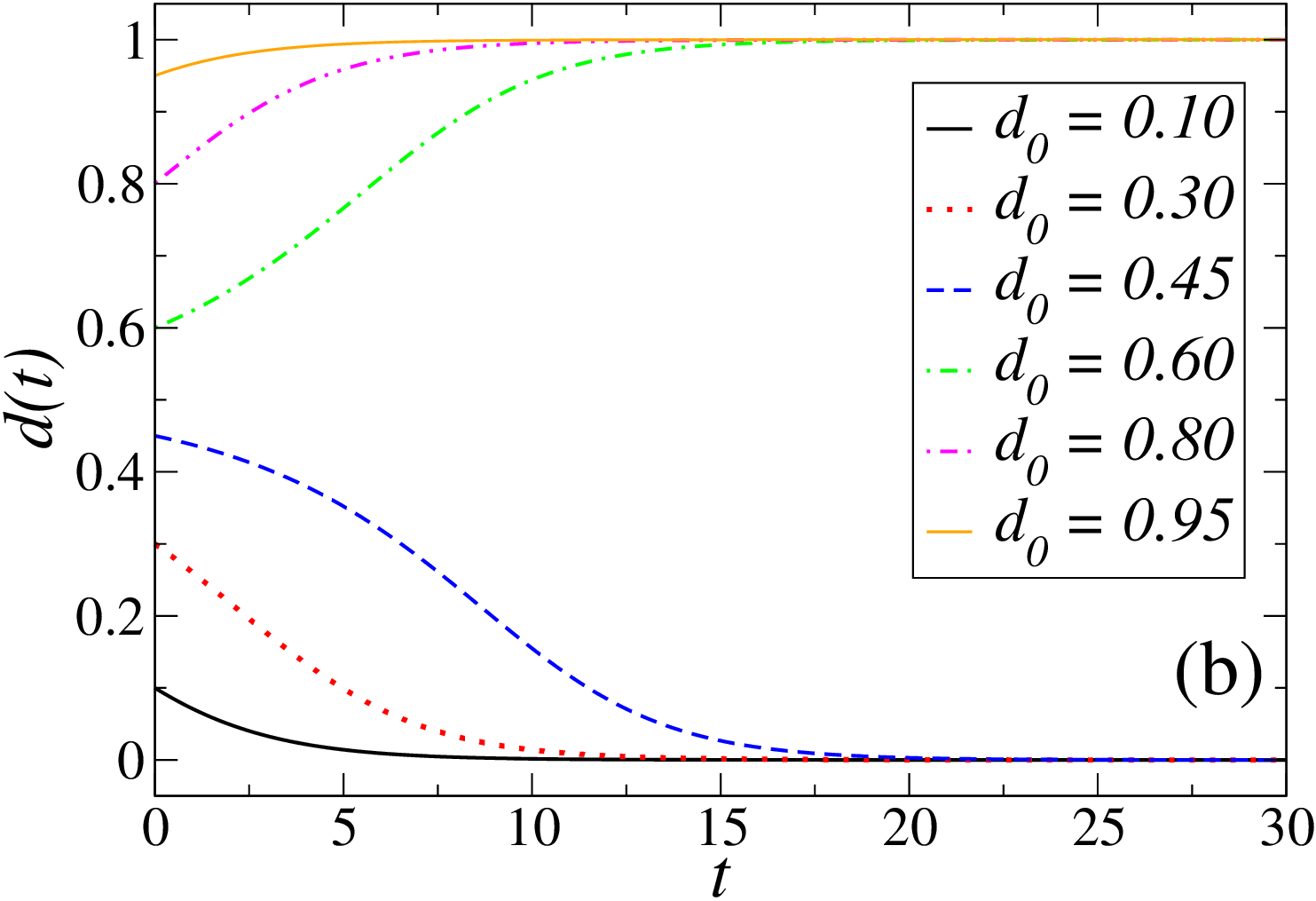}
\\
\vspace{1.0cm}
\includegraphics[width=0.35\textwidth,angle=0]{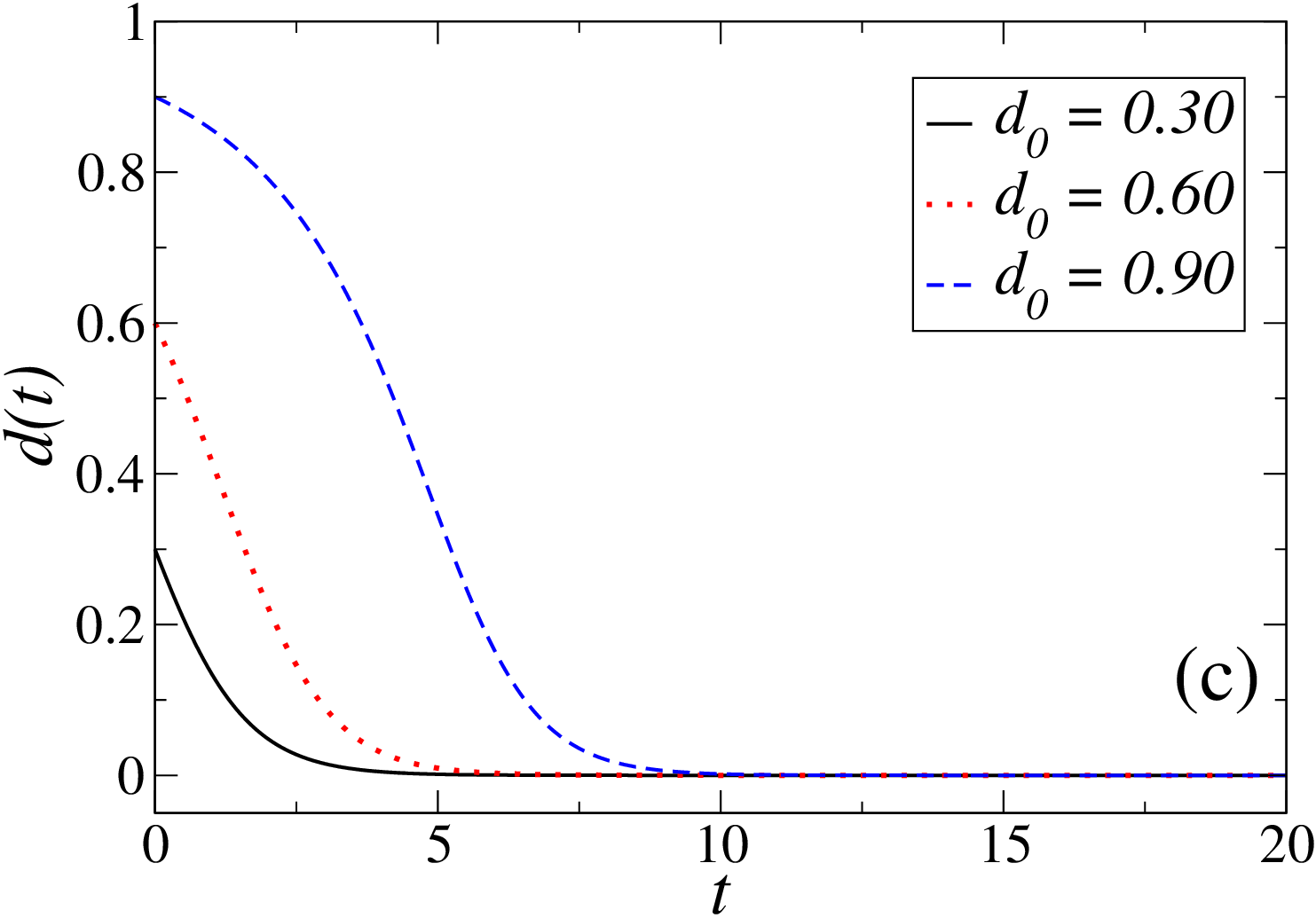}
\end{center}
\caption{Time evolution of the fraction $d(t)$ of rule-breakers for Model I for different initial conditions $d_0=d(t=0)$ and representative values of the punishment probability $p$. Panels (a)-(c) illustrate the three possible dynamical regimes of the Model I: (a) $p=0.03$, corresponding to $K>L$, where the population converges to full rule-breaking ($d=1$); (b) $p=0.15$, corresponding to $0<K<L$, where the system exhibits bistability depending on the initial condition; (c) $p=0.30$, corresponding to $K<0$, where the population converges to full compliance ($d=0$). The parameters are $B=1.0$, $F=5.0$, $C=0.2$, $s=0.3$, and $R=0.6$.}
\label{fig1}
\end{figure}

Without loss of generality, we will consider in this work $\gamma=1$. In order to illustrate the mentioned stable branches $d^*=1$ and $d^*=0$ and the bistable region, we exhibit in Fig. \ref{fig1} the time evolution of the fraction $d(t)$ of rule-breakers for Model I for different initial conditions $d_0=d(t=0)$ and typical values of the punishment probability $p$. The curves were obtained from the numerical integration of Eq. \eqref{eq6}. The parameters are $B=1.0$, $F=5.0$, $C=0.2$, $s=0.3$ and $R=0.6$. In Fig. \ref{fig1} (a) we show results for $p=0.03$, corresponding to the case $K>L$, where the population converges to full rule-breaking ($d=1$). We can see that, for the three initial conditions $d_0 = 0.30, 0.60$ and $0.90$, the population converges to $d=1$. In panel (b) the bistable region is clearly observed for $p=0.15$, where we have $0 < K < L$. For such value of $p$, Eq. \eqref{eq7} gives us $d^*=0.50$. Taking into account different initial conditions, we see that for $d_0<d^*=0.50$ the curves converge to $d=0$, whereas for $d_0>d^*=0.50$ they converge to $d=1$. Finally, in panel (c) we show results for $p=0.30$, corresponding to the case $K<0$, where the population converges to full compliance ($d=0$). The bistable region will be discussed in more details in the following.

From Eq. \eqref{eq7}, one can directly assess how the model parameters affect the system dynamics. In the bistable regime, the internal fixed point $d^*=1-K/L$ acts as a critical threshold separating the basins of attraction of full compliance ($d=0$) and widespread rule-breaking ($d=1$). Hence, increasing $d^*$ makes the rule-breaking regime harder to reach from small initial fractions of violators. This can be achieved by increasing $p$ and/or $F$, strengthening social sanctions by increasing $s$ and $R$, and reducing the private benefit $B$ or the compliance cost $C$.

From Eq. \eqref{eq7} we can obtain two threshold values for $p$, separating the stability of the branches $d^*=1$ and $d^*=0$. Taking $d^*=0$ and $d^*=1$ in Eq. \eqref{eq7}, we obtain the mentioned two values for $p$, namely
\begin{eqnarray} \label{eq8a}
p_1 & = & \frac{B + C - (s+R)}{F} \\ \label{eq9a}
p_2 & = & \frac{B+C}{F}
\end{eqnarray}

These expressions make explicit that the transition thresholds depend only on dimensionless parameter combinations, reinforcing that the qualitative behavior of the system is governed by relative incentive scales.

\begin{figure}[t]
\begin{center}
\vspace{6mm}
\includegraphics[width=0.3\textwidth,angle=0]{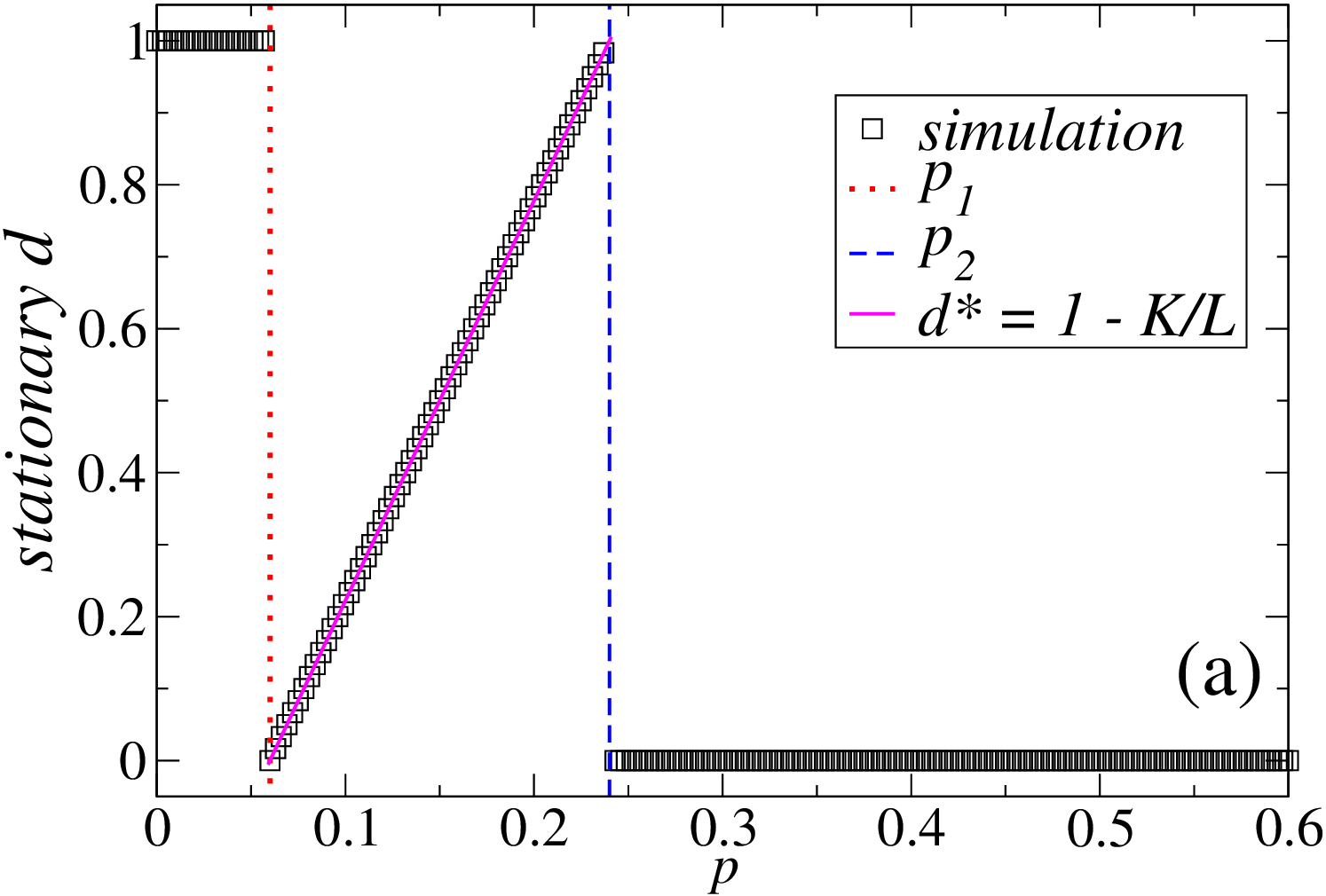}
\hspace{0.3cm}
\includegraphics[width=0.3\textwidth,angle=0]{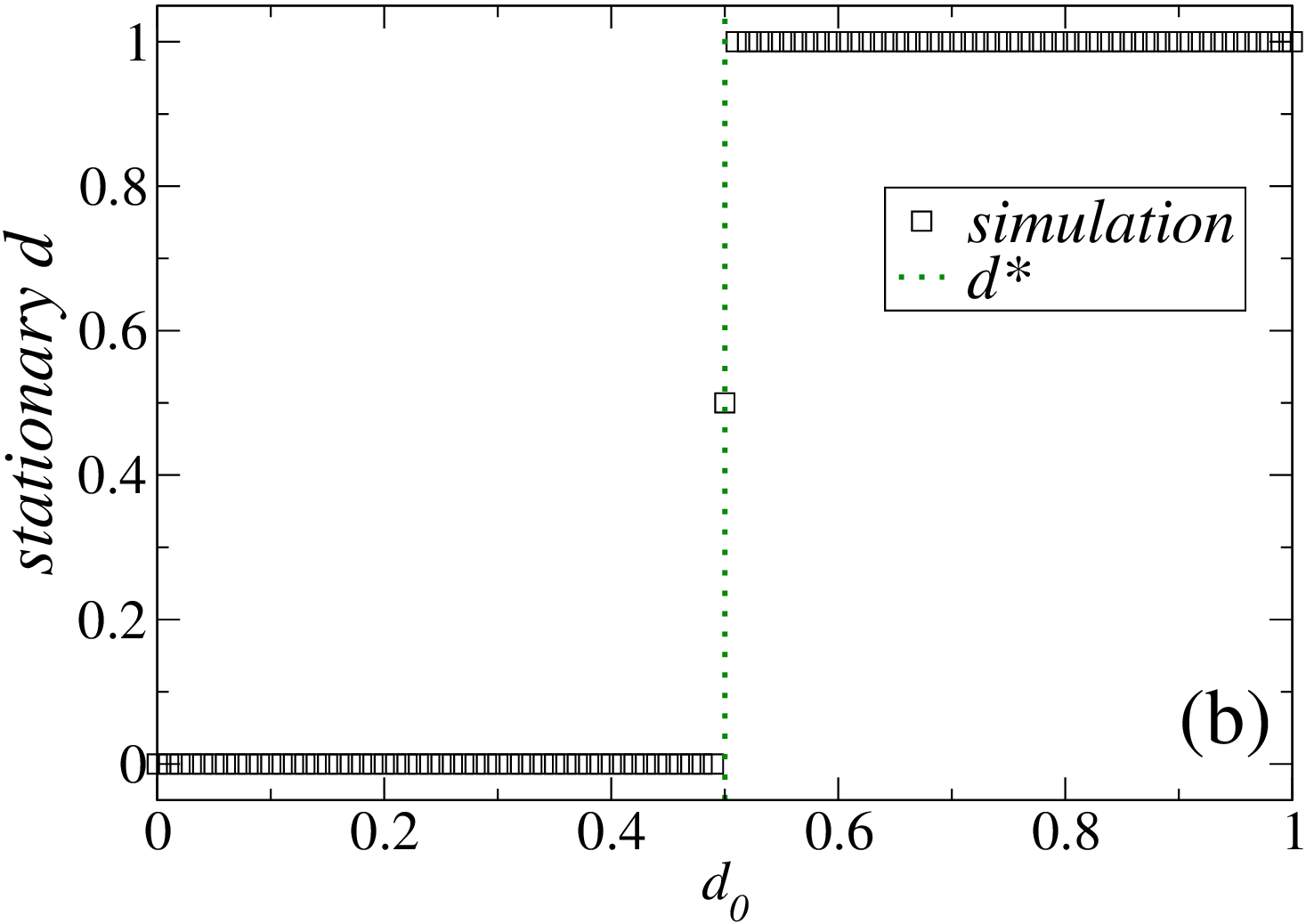}
\\
\vspace{0.5cm}
\includegraphics[width=0.3\textwidth,angle=0]{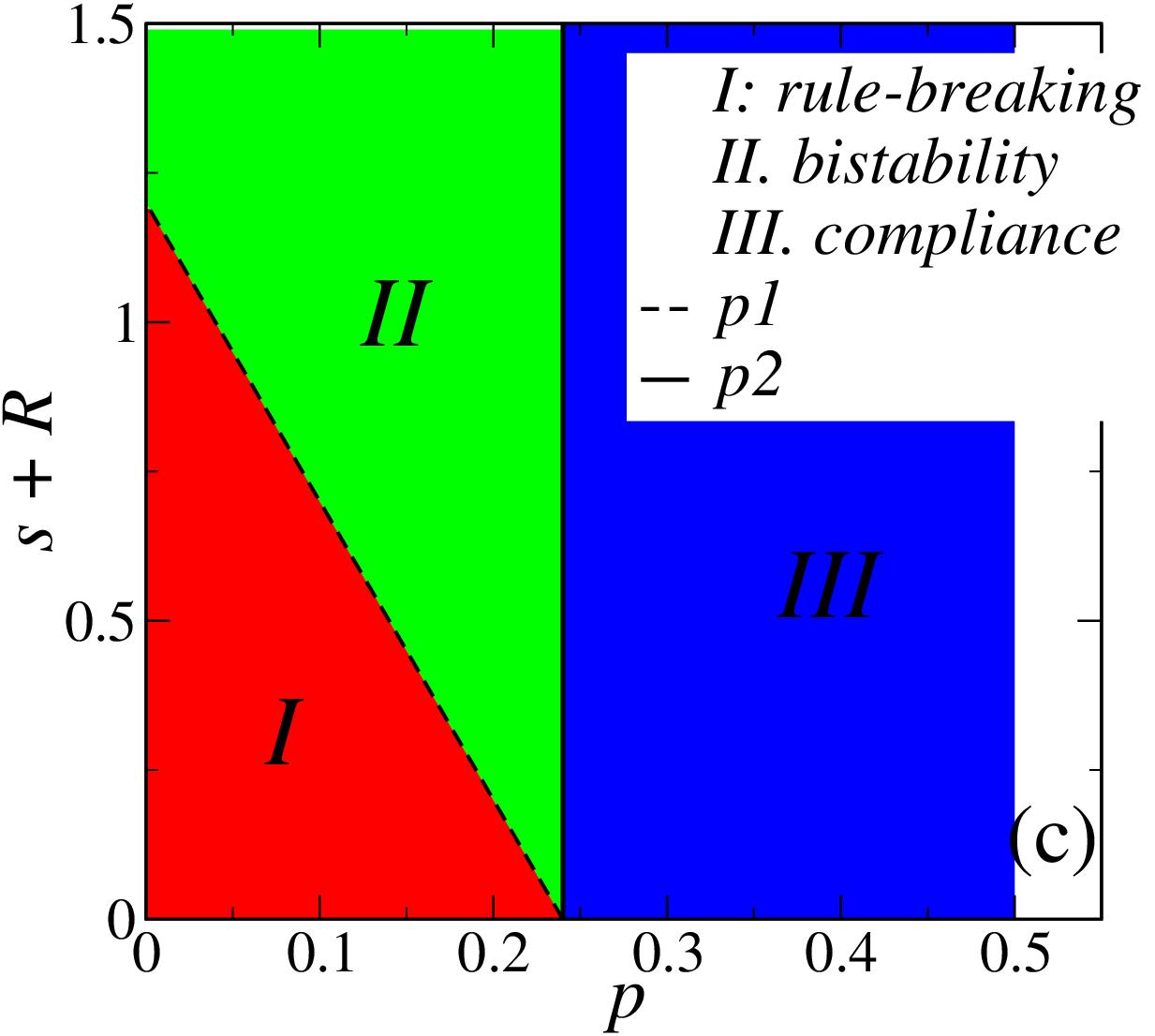}
\end{center}
\caption{Stationary behavior of Model I. (a) Stationary fraction of rule-breakers $d$ as a function of the punishment probability $p$. Numerical results (squares) show the stable branches $d=1$ and $d=0$, and the bistable region for $p_1 < p < p_2$, where $p_1=0.06$ and $p_2=0.24$ (Eqs. \eqref{eq8a} and \eqref{eq9a}) are indicated by dotted and dashed lines. The analytical solution $d^* = 1 - K/L$ (solid line) corresponds to the unstable intermediate branch. (b) Basin of attraction for $p=0.15$, showing the stationary value of $d$ as a function of the initial condition $d_0$. The threshold $d^* = 0.5$ separates the basins: $d_0 < d^*$ leads to $d=0$, while $d_0 > d^*$ leads to $d=1$. Parameters: $B=1.0$, $F=5.0$, $C=0.2$, $s=0.3$ and $R=0.6$. (c) Phase diagram in the $(s+R,p)$ plane. Regions correspond to full compliance ($d=0$, blue), bistability (green) and full rule-breaking ($d=1$, red). The boundaries are given by the analytical thresholds $p_1$ and $p_2$.}
\label{fig2}
\end{figure}

Thus, for $p<p_1$ the stable solution is $d^*=1$, whereas for $p>p_2$ the stable solution becomes $d^*=0$. For $p_1 < p < p_2$ the system exhibits a bistable regime in which both absorbing states are stable. In this interval, Eq. \eqref{eq7} defines the intermediate fixed point separating the two basins of attraction. An illustration of these threshold values is shown in Fig. \ref{fig2}, where the results were obtained from the numerical integration of Eq. \eqref{eq6}. In panel (a) we present the stationary fraction of rule-breakers $d$ as a function of the punishment probability $p$, for the same parameters considered in Fig. \ref{fig1}. For these parameters we obtain $p_1=0.06$ and $p_2=0.24$. The unstable branch given by Eq. \eqref{eq7} is observed in the interval $p_1 < p < p_2$, separating the two stable branches $d=0$ and $d=1$. It is worth noting that the width of the bistable region is given by $p_2 - p_1 = (s+R)/F$, showing that the coexistence of the two collective regimes is controlled by the strength of social feedback relative to institutional punishment. In panel (b) we show  the basin of attraction for $p=0.15$, showing the stationary value of $d$ as a function of the initial condition $d_0=d(t=0)$. The critical threshold $d^* = 1 - K/L = 0.5$ separates the two basins of attraction: for $d_0 < d^*$ the system converges to $d=0$, whereas for $d_0 > d^*$ it converges to $d=1$. The analytical result of Eq. \eqref{eq7}, which gives us $d^*=0.50$, is also shown for comparison (see the dotted line). This dependence on initial conditions reflects the presence of distinct basins of attraction in the bistable regime, indicating that different initial distributions of behavior may lead to qualitatively different collective outcomes. To further illustrate the structure of Model I, Fig. \ref{fig2}(c) shows the phase diagram in the $(s+R,p)$ plane, where the bistable region is identified between the analytical thresholds $p_1$ and $p_2$. This representation highlights how the interplay between enforcement ($p$) and social feedback ($s+R$) controls the emergence of discontinuous transitions and path dependence. The phase diagram also shows that the extent of the bistable region is highly sensitive to variations in the model parameters, particularly those controlling the balance between individual incentives and social feedback.

These results reinforce the idea that collective norm stability is governed by systemic incentives rather than by individual intentions, a hallmark of complex social systems. What appears to be an individual decision, such as violating a rule in a seemingly isolated context, is, in fact, embedded in a collective feedback loop. As the fraction of violators increases, the perceived utility of disobedience rises, pushing the system toward the threshold separating the basins of attraction and, ultimately, into a distinct macroscopic phase. 

The bistable dynamics described by Model I presents conceptual similarities with threshold models of collective behavior, such as the classical framework proposed by Granovetter \cite{Granovetter}. In these models, individuals adopt a behavior once the fraction of adopters in the population exceeds a personal threshold. In the present formulation, the internal fixed point $d^{*}$ plays an analogous role at the macroscopic level, separating the basins of attraction of widespread compliance and generalized rule-breaking. In this sense, abrupt transitions toward large-scale noncompliance may be interpreted as collective cascades triggered when the system crosses this critical threshold.

The bistable regime predicted by Model I also resembles hysteresis phenomena widely studied in socioeconomic systems, where societies may become trapped in alternative stable equilibria depending on their historical trajectory \cite{rmp,Schelling2,Bicchieri}. In such situations, transitions between high-compliance and widespread rule-breaking regimes require sufficiently large perturbations to overcome the collective threshold separating the two basins of attraction.

Given the one-dimensional nature of the dynamics, it is also possible to introduce an effective potential function $V(d)$ such that $\dot d = - \frac{dV}{dd}$. In this representation, the stable stationary states correspond to local minima of $V(d)$, while the unstable fixed point defines a maximum separating the basins of attraction. Although this representation is not required for the present analysis, it provides an alternative interpretation of the bistable dynamics observed in Model I.


\subsection{Model II: Negative social feedback and continuous transition}

The expected utility of breaking the rule in Model II is written as
\begin{equation} \label{eq8}
U_D(d) = B - pF - \alpha\,d,
\end{equation}
\noindent
where we preserve a similar structure introduced in Model I, but now account for an increasing collective cost of rule-breaking through a term proportional to the fraction $d$ of disobedient agents, controlled by a parameter $\alpha$. The parameters $B$, $p$ and $F$ retain the same meaning as in subsection 2.2.

The expected utility of compliance is given by
\begin{equation} \label{eq9}
U_C(d) = -C + R ,
\end{equation}
\noindent
where $C$ and $R$ have the same interpretation as in subsection 2.2.

In contrast to Model I, the linear penalization term proportional to $d$ introduces a stabilizing feedback, leading to a continuous transition between compliance and disobedience, as we will see in the following.

The utility difference can then be written as
\begin{equation} \label{eq10}
\Delta^{'}(d) = U_D(d) - U_C(d) = (B - pF + C - R) - \alpha\,d = K^{'} - L^{'}\,d
\end{equation}
with
\begin{equation} \label{eq11}
K^{'} = B - pF + C - R, \qquad L^{'} = \alpha.
\end{equation}

The system also follows a mean-field dynamics,
\begin{equation} \label{eq12}
\dot d(t) = \gamma\, d(1-d)\,\Delta^{'}(d) ~.
\end{equation}
\noindent
Again, the absorbing states $d=0$ and $d=1$ are also fixed points of the dynamics. However, now for $0 < K^{'} < L^{'}$ we have a stable fixed point given by
\begin{equation} \label{eq13}
d^* = \frac{K^{'}}{L^{'}} = \frac{B - pF + C - R}{\alpha} ~.
\end{equation}
\noindent
In contrast to Model I, the linear penalization term proportional to $d$ introduces a stabilizing social feedback: as the fraction of disobedient agents increases, the relative advantage of rule-breaking is progressively reduced. As a consequence, the internal fixed point $d^* = K'/L'$ becomes stable whenever $0 < K' < L'$, leading to a gradual and continuous change in the stationary value of $d$. This behavior is characteristic of continuous phase transitions, in which the order parameter varies smoothly as control parameters are tuned, and no bistability or hysteresis is observed.

It is important to emphasize the conceptual difference between $d^*$ in Models I and II. In Model I, the internal fixed point $d^*$ acts as a critical threshold separating the basins of attraction of full compliance and widespread rule-breaking. In contrast, in Model II, $d^*$ corresponds to the unique stationary fraction of rule-breakers.

\begin{figure}[t]
\begin{center}
\vspace{6mm}
\includegraphics[width=0.35\textwidth,angle=0]{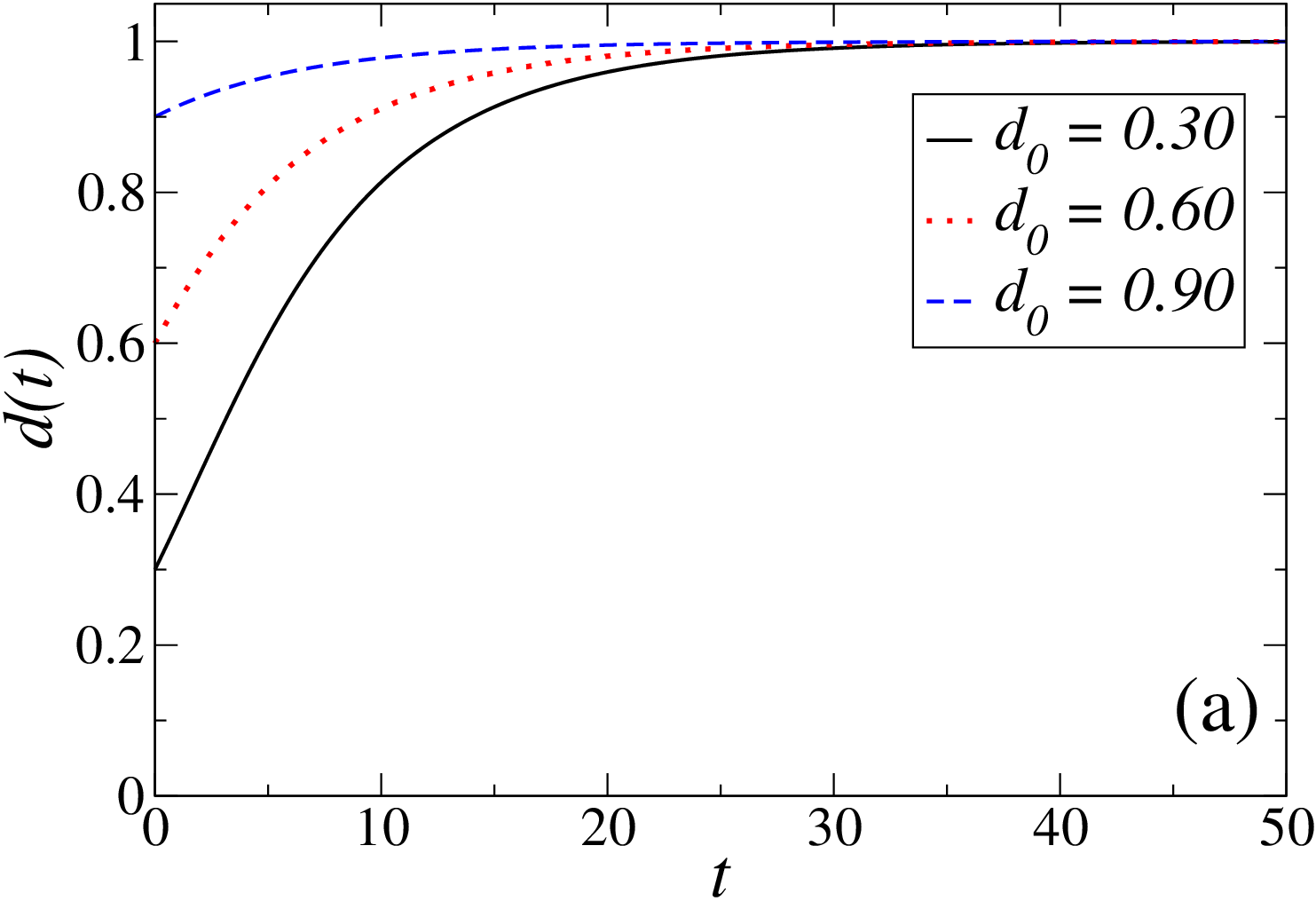}
\hspace{0.3cm}
\includegraphics[width=0.35\textwidth,angle=0]{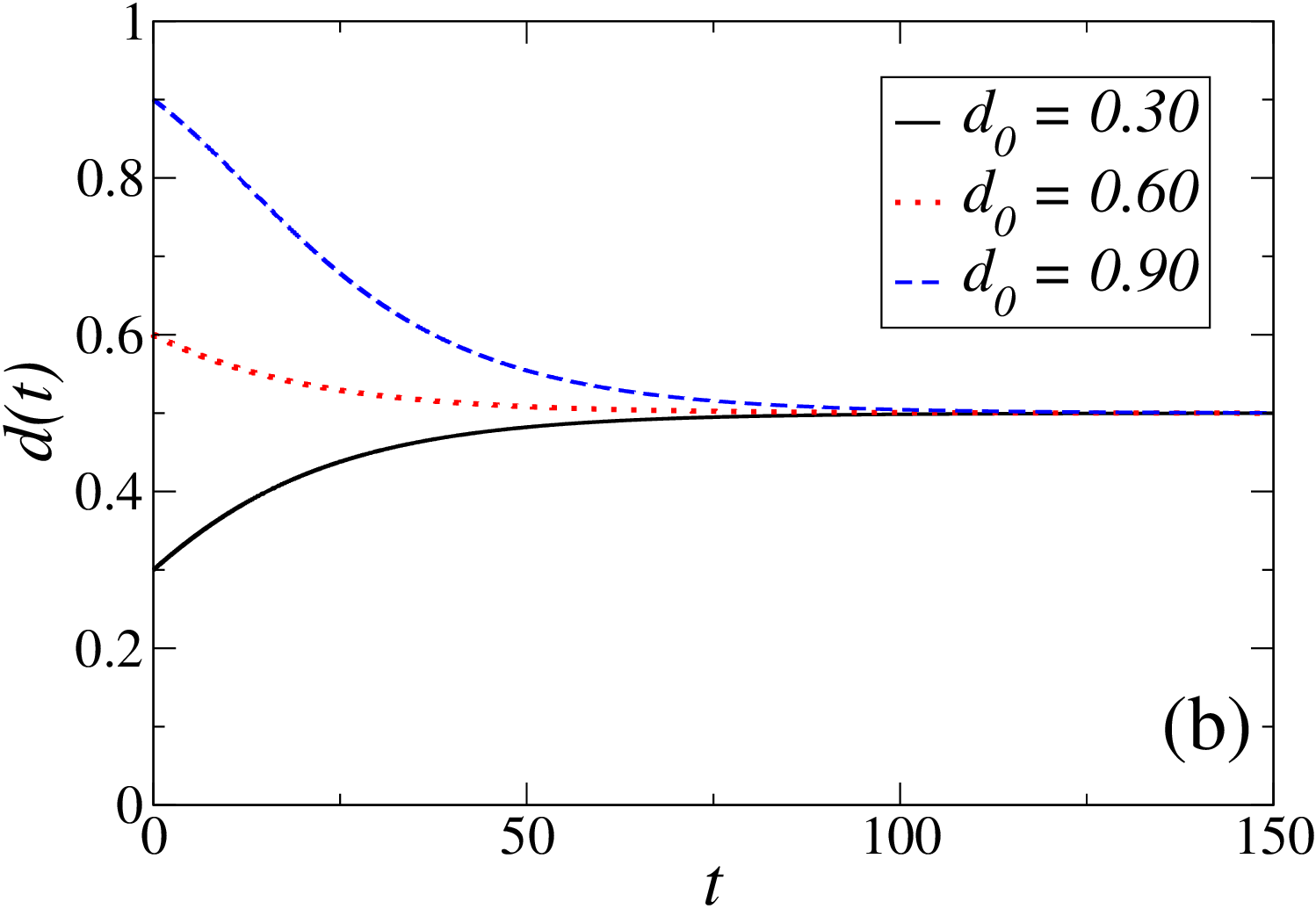}
\\
\vspace{1.0cm}
\includegraphics[width=0.35\textwidth,angle=0]{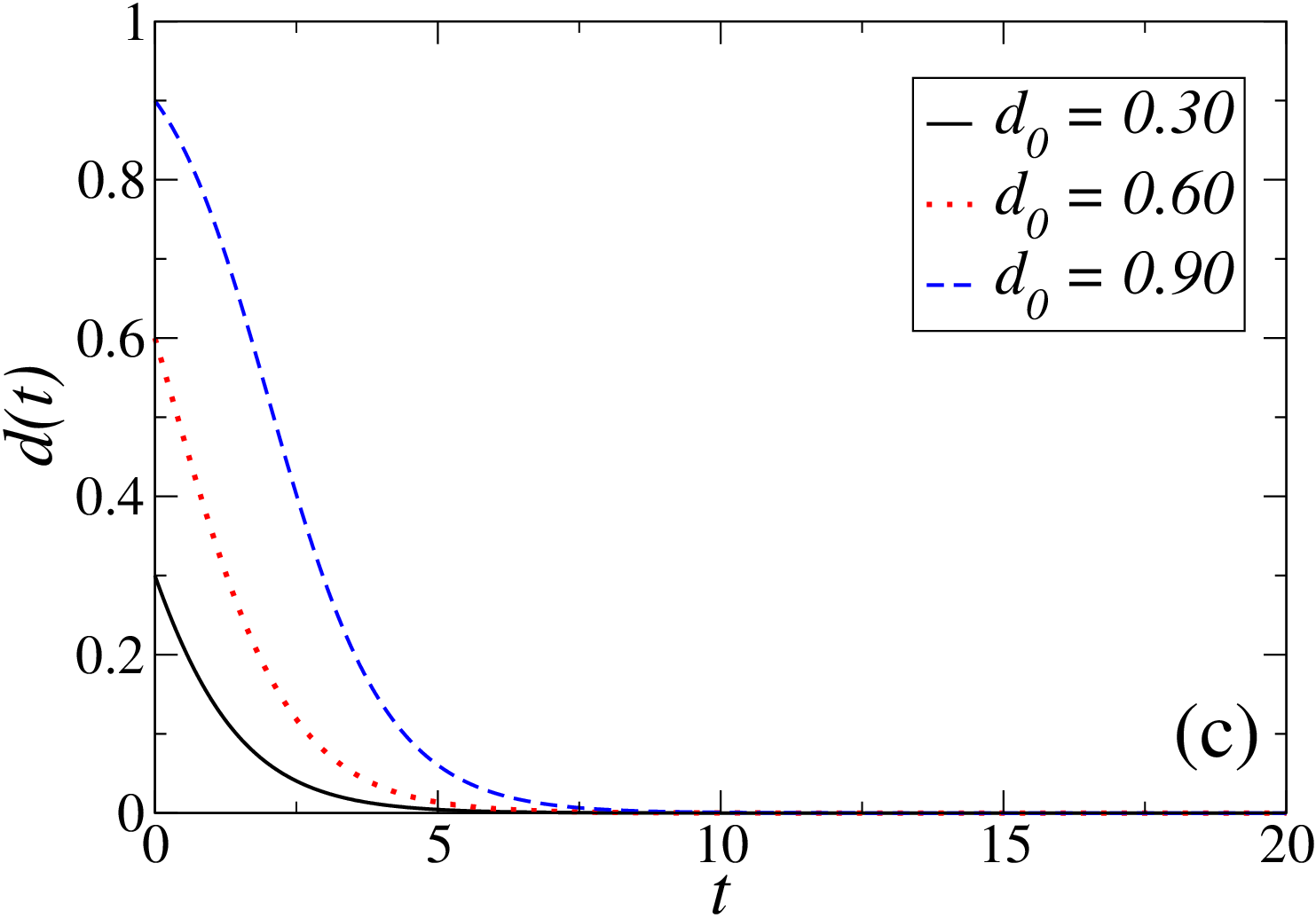}
\end{center}
\caption{Time evolution of the fraction $d(t)$ of rule-breakers for Model II for different initial conditions $d_0=d(t=0)$ and representative values of the punishment probability $p$. The parameters are $B=1.0$, $F=5.0$, $C=0.2$, $R=0.6$ and $\alpha=0.20$, which leads to $p_c=0.12$ obtained from Eq. \eqref{eq14}. Panel (a) shows results for $p=0.05$, where the population converges to full rule-breaking ($d=1$), panel (b) exhibits results for $p=0.10$, where the population reaches the stationary state $d=0.50$ given by Eq. \eqref{eq13}, for any initial condition, and panel (c) shows results for $p=0.30$,  where the population converges to full compliance ($d=0$).}
\label{fig3}
\end{figure}

The difference between Models I and II can become clearer when we observe the time evolution of the fraction of rule-breakers in Model II. In Fig. \ref{fig3} we exhibit numerical results for $d(t)$ for Model II for different initial conditions $d_0=d(t=0)$ and representative values of the punishment probability $p$. The parameters are $B=1.0$, $F=5.0$, $C=0.2$, $R=0.6$ and $\alpha=0.20$. Panel (a) shows results for $p=0.05$, showing the convergence of the population to full rule-breaking ($d=1$). In panel (b) it is shown results for $p=0.10$, where we can see the convergence for a stationary value $0<d^*<1$ of rule breakers, specifically we have $d^*=0.50$ from Eq. \eqref{eq13}.  Finally, panel (c) exhibits results for $p=0.30$, an illustration of the convergence to full compliance ($d=0$). These results will be better understood in the following. 

Results in Fig. \ref{fig3} suggest a continuous active-absorbing phase transition between rule-breaking and rule compliance in Model II. Indeed,  Eq. \eqref{eq13} can be rewritten as $d^* \sim (p_c-p)^{\beta}$, where the critical exponent is $\beta=1$ and the critical point is given by
\begin{equation} \label{eq14}
p_c = \frac{B + C - R}{F}    
\end{equation}
\noindent
This linear behavior corresponds to a mean-field critical exponent $\beta = 1$, consistent with the absence of spatial correlations in the present formulation \cite{Dickman}.

\begin{figure}[t]
\begin{center}
\vspace{6mm}
\includegraphics[width=0.5\textwidth,angle=0]{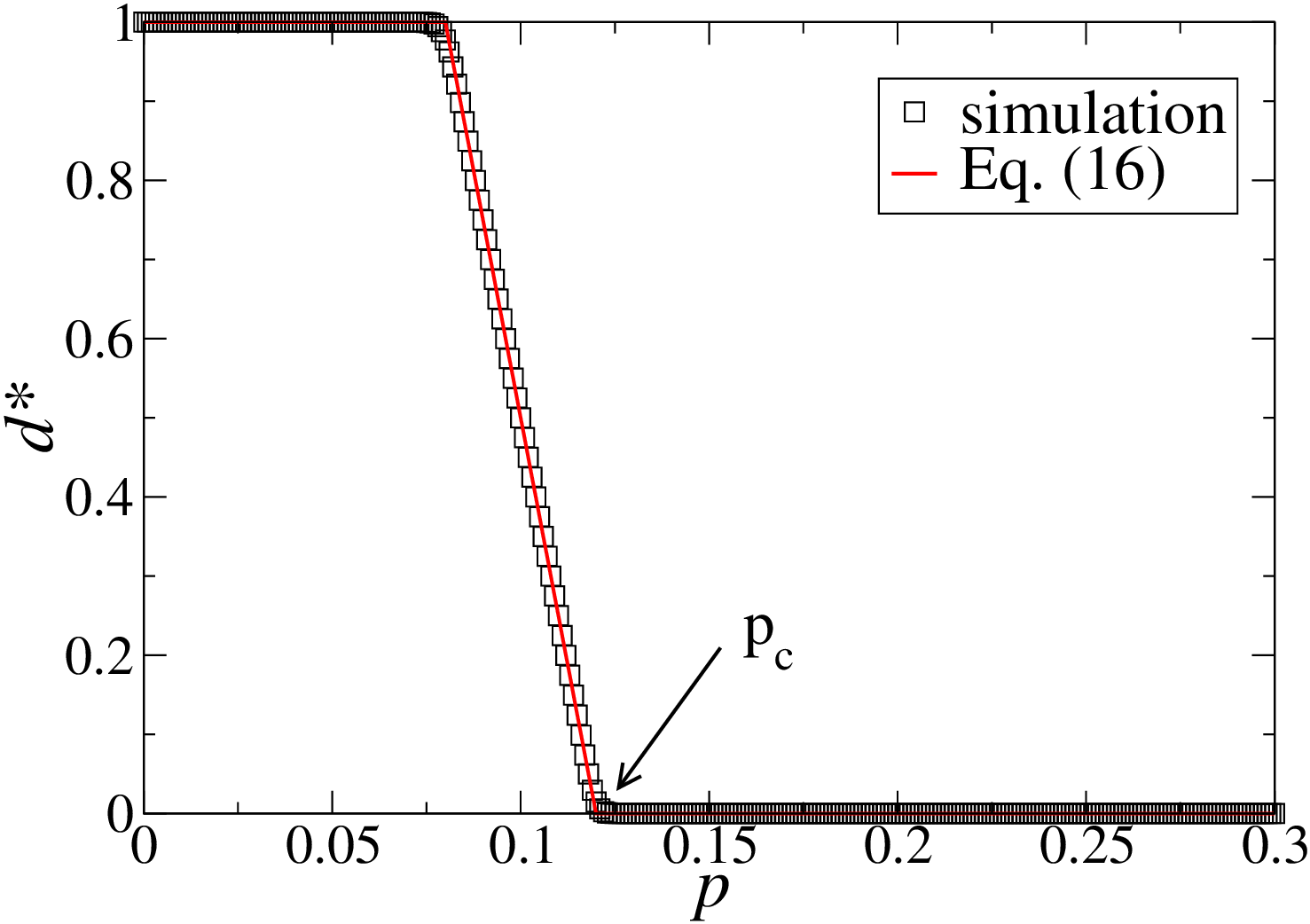}
\end{center}
\caption{Stationary fraction of rule-breakers $d^{*}$ for Model II as a function of the punishment probability $p$, obtained from the numerical integration of Eq. \eqref{eq12} (squares). The analytical result of Eq. \eqref{eq13} is also show for comparison (full line). The parameters are $B=1.0$, $F=5.0$, $C=0.2$, $R=0.6$ and $\alpha=0.20$, which leads to $p_c=0.12$ obtained from Eq. \eqref{eq13}, indicated by an arrow. The continuous nature of the phase transition is clearly observed.}
\label{fig4}
\end{figure}

In Model II, the stationary fraction of rule-breakers is given by $d^* = K'/L' = (B - pF + C - R)/\alpha$, which allows a direct assessment of parameter effects. Unlike Model I, where the internal fixed point plays the role of a threshold, here $d^*$ corresponds to the macroscopic stationary fraction of rule-breaking. Increasing the probability or severity of punishment ($p$ or $F$), strengthening reputational rewards $R$ or enhancing the collective cost of violations $\alpha$ reduces $d^*$. Conversely, higher private benefits of disobedience $B$ or larger compliance costs $C$ increase the stationary fraction of rule-breakers. An illustration of the continuous transition in Model II is given in Fig. \ref{fig4}, where analytical and numerical results are exhibited for the stationary fraction of rule-breakers $d^*$ as a function of the punishment probability $p$.

From the viewpoint of statistical physics, the continuous transition described in Model II belongs to the broader class of nonequilibrium collective transitions in interacting populations, where macroscopic behavioral states emerge from simple microscopic incentives and feedback mechanisms \cite{Dickman}.


\subsection{Everyday manifestations of the collective dynamics}

The examples discussed below are not intended as specific case studies, but rather as illustrations of how everyday social situations may fall into different dynamical regimes depending on the structure of collective feedback. In all cases, the same population of individuals may display either abrupt norm breakdown or stable mixed behavioral patterns, as captured by Models I and II, respectively. Whenever appropriate, we explicitly refer to the model parameters in order to connect these situations to the theoretical framework.

\textbf{(i) Traffic violations:}
\\

Traffic-related rule-breaking provides a paradigmatic example of collective behavioral dynamics. When enforcement is weak (low $pF$) and individual benefits $B$ are high, violations may spread through reinforcing social mechanisms, leading to widespread noncompliance, as described by Model I. As the fraction of violators $d$ increases, however, congestion and conflicts reduce the effective benefits of violations, introducing a collective cost captured by $\alpha$. This stabilizing feedback leads to intermediate levels of compliance, consistent with Model II.

\vspace{0.3cm}

\noindent
\textbf{(ii) Academic dishonesty:}
\\

Academic misconduct illustrates rule-breaking driven by individual incentives and collective feedback. When supervision is weak ($p \approx 0$) and the benefit $B$ is significant, dishonest behavior may spread through positive social reinforcement, eventually leading to widespread noncompliance, in agreement with Model I. As misconduct becomes more frequent, institutions typically introduce monitoring technologies and integrity policies, increasing the effective cost of violations $\alpha$ and enhancing the reward $R$ for compliance. These mechanisms promote a gradual transition toward stable intermediate levels of compliance, as described by Model II.

\vspace{0.3cm}

\noindent
\textbf{(iii) Tax compliance:}
\\

Tax evasion provides another example of the interplay between private incentives and collective regulation. When enforcement is weak ($p \approx 0$) and the benefit $B$ is high, evasion may spread through reinforcing mechanisms, driving the system toward widespread noncompliance (Model I). By contrast, improved monitoring (higher $pF$), reduced compliance costs $C$, and increased reputational incentives $R$ reshape the collective incentives, leading to stable intermediate levels of compliance through a continuous transition, as described by Model II.

\vspace{0.3cm}

\noindent

Beyond these examples, similar mechanisms may operate in a wide range of social contexts, including informal norm violations in shared spaces, fare evasion and digital piracy. Although these situations differ in scale and institutional setting, they share the same structural features of individual incentives coupled to collective feedback, and can therefore be interpreted within the same modeling framework.

Together, these examples illustrate that the qualitative nature of collective outcomes is not determined by individual intentions or moral judgments, but by the feedback structures shaping incentives at the population level. Consequently, the same everyday situation may display distinct dynamical behaviors depending on whether collective interactions amplify or penalize widespread rule-breaking.


\section{Final Remarks}

In this work, we have proposed a minimal dynamical framework to describe the collective persistence of rule-breaking behavior in social systems. Although the individual-level assumptions are simple and based on utility comparisons, the resulting macroscopic dynamics reveal two structurally distinct classes of collective transitions.

When social reinforcement dominates, as described in Model I, the system exhibits bistability and a discontinuous transition between full compliance and widespread rule-breaking. In this regime, the internal fixed point acts as a critical threshold separating basins of attraction. Small variations in parameters or initial conditions may therefore trigger abrupt norm collapse, and the system may display strong path dependence. Such behavior resembles first-order phase transitions in statistical physics, where coexistence and hysteresis naturally emerge from reinforcing interactions.

In contrast, Model II captures situations in which the collective cost of widespread violations increases with their frequency. In this case, the transition between compliance and rule-breaking becomes continuous, and the stationary fraction of violators varies smoothly with model parameters. Rather than abrupt collapse, the system settles into intermediate levels of compliance, reflecting a stabilizing feedback mechanism. This behavior is analogous to continuous phase transitions, in which macroscopic order changes gradually as control parameters vary.

The mechanisms proposed in the \textit{Broken Windows theory} can also be interpreted within the framework developed here. The Broken Windows theory, originally proposed by Wilson and Kelling \cite{broke_windows}, suggests that visible signs of minor disorder or norm violations may encourage further violations by signaling weak social or institutional control. Field experiments by Keizer et al. \cite{Keizer} provide empirical support for this mechanism, showing that visible disorder such as graffiti or littering substantially increases the probability of additional norm violations. Importantly, the observed effect is typically an increase in violation rates rather than a transition toward universal noncompliance. This empirical feature is naturally captured by the stabilizing dynamics of Model II: disorder cues may increase the propensity to violate rules, but as violations become more frequent the associated collective costs grow, introducing negative feedback (encoded by the parameter $\alpha$) that prevents a complete breakdown of norms and stabilizes intermediate stationary levels of compliance. By contrast, if disorder cues primarily reinforce the perception that rules are weakly enforced or socially tolerated, the reinforcing feedback may dominate. In such situations, even small initial levels of rule-breaking can trigger rapid collective shifts toward widespread noncompliance, driving the system toward the bistable regime described by Model I.

From an empirical perspective, the proposed framework may be tested by comparing its qualitative predictions with time series of collective behavior in real systems, such as the fraction of individuals violating specific norms. In such cases, model parameters could be inferred by matching observed transitions between regimes of compliance and noncompliance. From a policy perspective, the model also suggests mechanisms to prevent abrupt transitions toward widespread rule-breaking. In particular, increasing the effective enforcement (through higher $pF$), strengthening social norms (larger $s$ or $R$), or enhancing the collective cost of violations ($\alpha$) may shift the system away from bistable regions, thereby reducing the likelihood of sudden transitions and promoting stable compliance regimes.

The present formulation is intentionally minimal. It assumes homogeneous agents, mean-field interactions and static parameters. Extensions may include heterogeneous agents, explicit network structures, stochastic fluctuations or time-dependent institutional responses. Such generalizations may allow a closer connection with empirical data and cross-cultural comparisons.

In particular, the inclusion of network structure may influence the quantitative location of transition points and the size of bistable regions. Heterogeneous connectivity patterns and local clustering can modify the effective feedback experienced by individuals, potentially smoothing abrupt transitions or enhancing local cascades. However, the qualitative distinction identified in this work between reinforcing feedback leading to bistability and stabilizing feedback leading to continuous transitions is expected to remain robust, as it is rooted in the functional form of the feedback mechanisms rather than in the specific interaction topology.

The present analysis is based on a deterministic mean-field formulation, which does not include stochastic fluctuations or finite-size effects. In more realistic settings, such effects may influence the behavior near transition points, for example through fluctuations around the stationary states or finite-size rounding of discontinuities. Incorporating such effects would require stochastic or network-based extensions of the model.

The distinction between the two models does not arise from differences in individual rationality, but from the structural nature of collective feedback. In this sense, the main contribution of the present work is to show that different feedback mechanisms lead to qualitatively distinct classes of collective behavior, determining whether social transitions occur abruptly or continuously. This provides a unified framework linking the structure of social interactions to the order of social phase transitions.

More broadly, the framework illustrates how tools from statistical physics and evolutionary dynamics can contribute to the understanding of social order and its fragility. Collective norm stability, much like physical order, depends on the balance between reinforcing and stabilizing interactions. Identifying which feedback mechanisms dominate in a given context may therefore be central to understanding when societies experience abrupt norm collapse or gradual behavioral adaptation.

In this perspective, the persistence or breakdown of social norms is not merely a matter of individual morality, but a dynamical property of collective systems governed by feedback structure.


\section*{Acknowledgments}

The author acknowledges partial financial support from the Brazilian scientific funding agency Conselho Nacional de Desenvolvimento Científico e Tecnológico, Brazil (CNPq, Grants 308643/2023-2 and 406820/2025-2). I am grateful to S. Goulart for bringing the Broken Windows theory to my attention and suggesting its possible relation to the model.

\bibliographystyle{elsarticle-num-names}

\end{document}